\documentclass[]{revtex4-1}

\usepackage{amsmath}
\usepackage{amsfonts}
\usepackage{amsthm}
\usepackage{amssymb}
\usepackage{graphicx}
\usepackage{hyperref}

	\newcommand{\ncd}{\newcommand}
	\ncd{\mrm}    {\mathrm}
	\ncd{\beq} {\begin{equation}}
	\ncd{\eeq} {\end{equation}}

	\def\d{{\rm d}}
	
	\def\tGamma{\tilde\Gamma}

	\newtheorem{theo}{Theorem}
	
\begin{document}

	%\title{Fluctuation-dissipation  through Geometric Thermalisation}
	\title{Para-Sasakian geometry in thermodynamic fluctuation theory}

	\author{A. Bravetti}
        \email{bravetti@correo.nucleares.unam.mx}
	\affiliation{Instituto de Ciencias Nucleares, Universidad Nacional Aut\'onoma de M\'exico,\\ AP 70543, M\'exico, DF 04510, Mexico.}

	\author{C. S. Lopez-Monsalvo}
	\email{cesar.slm@correo.nucleares.unam.mx}
	\affiliation{Instituto de Ciencias Nucleares, Universidad Nacional Aut\'onoma de M\'exico,\\ AP 70543, M\'exico, DF 04510, Mexico.}

	\begin{abstract}
In this work we tie concepts derived from statistical mechanics, information theory and contact Riemannian geometry within a single consistent formalism for thermodynamic fluctuation theory. We derive the concrete relations characterizing the geometry of the Thermodynamic Phase Space stemming  from the relative entropy and the Fisher-Rao information matrix. In particular, we show that the Thermodynamic Phase Space is endowed with a natural para-contact pseudo-Riemannian structure derived from a statistical moment expansion which is para-Sasaki and $\eta$-Einstein. Moreover, we prove that such manifold is locally isomorphic to the hyperbolic Heisenberg group. In this way we show that the hyperbolic geometry and the Heisenberg commutation relations on the phase space naturally emerge from classical statistical mechanics. Finally, we argue on the possible implications of our results.
 	\end{abstract}

\maketitle

\section{Introduction}

Information theory has been widely used in many branches of science, spanning systems from quantum mechanics to biology, from cosmology to statistical 
inference. In this context, particular attention has been devoted to the notion of the relative entropy - or Kullback-Leibler divergence \cite{KLD} - which 
gives an estimation of the information gain (or loss) that one realizes when passing from a probability distribution to another. 
%Such an entropy has a number of interesting properties related to both equilibrium
%and non-equilibrium thermodynamics. Moreover, the relative entropy has also been used in the context of the
%AdS/CFT correspondence (see e.g. \cite{BCHMyers2013}) to give a generalized statement of the Bekenstein bound which holds in Quantum Field Theory.
The relative entropy is a functional whose arguments are pairs of distribution functions. It has been shown {  in 
\cite{Ingarden} that 
%\cite{BCHMyers2013} that if one chooses one of the two distributions to be Gibbs equilibrium (canonical) distribution, then 
the first order variation of the relative entropy vanishes. In the case of statistical mechanics, this is the statement of  the first law of thermodynamics.}

Another fundamental concept in information theory is that of the Fisher-Rao Information Matrix, which provides us with another measure of the distance between two different probability distributions. Such a measure endows the statistical manifold with a Riemannian structure (see e.g. \cite{MNSS1990,Schlogl, BrodyRivier,BrodyRitz,BrodyHook}). In fact, while the relative entropy does not define a real distance between distributions (for example, it is not symmetric),  it can be shown that the Fisher-Rao Information Matrix arises as the Hessian of the relative entropy over a stationary point.  The entries of such a matrix are in correspondence with the components of the metric tensor over the manifold of probability distributions \cite{BrodyRivier,BrodyRitz,BrodyHook}. Furthermore, when restricted to Gibbs equilibrium probability distributions, the Fisher-Rao Information Metric has been used to define a thermodynamic length on the space of equilibrium states of thermodynamic systems. This length is equivalent (up to Legendre transformations) to other definitions given in the literature \cite{wein1975, rupp1979}, but it
has the advantage that it can be extended to the non-equilibrium case \cite{Schlogl, BrodyRivier,BrodyRitz,Crooks,CrooksPRE2012,CrooksPRL2012}. 
 One of the most important properties of the thermodynamic length is that, for paths out of equilibrium, it gives a bound to the {  loss of available work -- or \emph{dissipated availability} --} during the process \cite{SalamonBerryPRL}.  Moreover,
measurements of the thermodynamic length can be obtained from non-equilibrium protocols, i.e. this quantity influences also the behavior of systems out of equilibrium and hence can be used to obtain optimal paths (see \cite{CrooksPRE2012,CrooksPRL2012}).
%{  However, it should be mentioned that all these applications are within the assumption of local equilibrium.}

Interestingly, the construction of a Riemannian geometry over the manifold of equilibrium states of a thermodynamic system has been generalized in \cite{MNSS1990}. In such work, the authors showed that the statistical construction naturally endows both, the manifold of equilibrium states and the phase space of equilibrium thermodynamics, with a pseudo-Riemannian structure. In particular, the phase space turns out to be equipped with both a contact and a pseudo-Riemannian structures, whose restrictions to Legendre sub-manifolds define the equilibrium manifold itself and the thermodynamic version of the Fisher-Rao Information Metric, respectively.

%Among contact Riemannian manifolds, a prominent role is played by Sasakian manifolds. Sasakian manifolds are the odd-dimensional counterpart of K\"ahler manifolds and, as such, they share a lot of desirable properties. In particular, in recent years they have been the focus of attention for theoretical physics due to the fact that in the AdS/CFT duality, one generally deals with a product manifold $\mathrm{AdS}_{5}\times Y$, where $\mathrm{AdS}_{5}$ is the Anti-de-Sitter spacetime in five dimensions and $Y$ is a Sasaki-Einstein manifold. Thus, the gauge/gravity correspondence prescribes that most of the symmetries and hence the type of the dual field theory are encoded in the symmetries of the Sasakian manifold $Y$.  For example, it has been  shown in \cite{MartelliSparksYau} that the symmetries generated by the Reeb vector field of $Y$ correspond to the R-symmetries of the dual theory and that the volume of the Sasaki-Einstein manifold is directly related to the a central charge.

In this work,  we generalize the construction in \cite{MNSS1990} in order to show that the contact and pseudo-Riemannian structures
of the phase space can be defined also for distributions that are different from the Gibbs equilibrium distribution. This is important for extending this geometric picture to systems out of equilibrium. Moreover, in our treatment we will make clear that these structures can be derived elegantly by means of the variation of the relative entropy. Finally, we will show that the phase space of thermodynamics -- as defined in \cite{MNSS1990} --  possesses a number of intriguing  geometric properties. In particular, it is a para-Sasakian and $\eta$-Einstein manifold whose Ricci  scalar  of the  Levi-Civita connection
is constant {  (see the standard references \cite{SASAKI,Yano,bookBlair} for Sasakian manifolds and \cite{Takahashi,Zamkovoy,IVZ,Perrone} for the para-Sasakian case).}

In the context of para-Sasakian geometry there is another connection of  geometric significance which is parallel with respect to the metric and the other  tensors defining the contact-metric strucuture. We refer to this as the canonical connection \cite{Zamkovoy,IVZ}. {  The main result of this work is a proof that} the Thermodynamic Phase Space equipped with such  connection has vanishing  curvature.
This implies that the whole para-Sasakian structure is locally isomorphic to the hyperbolic Heisenberg group. A number of implications will be considered in the conclusions.

This paper is structured as follows. In Section \ref{sec2} we derive the Fisher-Rao Information Metric from two different perspectives, on the one hand we use  a statistical moment expansion of the differential entropy and, on the other hand, by means of the relative entropy. Then, in Section \ref{sec3},  we revisit the manner in which the construction of Section \ref{sec2} equips the Thermodynamic Phase Space with a contact-Riemannian structure. In Section \ref{sec4} we present new results regarding the algebraic and geometric structure of this space. In particular we show that this corresponds to the hyperbolic Heisenberg group. Finally, in Section \ref{sec5} we provide some closing remarks and comment on the implications of this construction in a more general setting.

\section{Two roads to the Fisher-Rao information metric}
\label{sec2}

In this Section we will revisit the work of Mruga{\l}a et al. \cite{MNSS1990}. In such work, the authors proved that the phase space of thermodynamics -- obtained after averaging out the phase space $\Gamma$ of statistical mechanics with respect to the optimal probability distribution $\rho_0$ --  is naturally endowed with both, a contact and a pseudo-Riemannian, structures arising from the maximization of the entropy functional.
Here, we propose a more general formulation which can be applied to any distribution (not necessarily the Gibbs one). Therefore, our approach is  relevant for generalizations to systems out of equilibrium, as we will argue. Moreover, we will derive the Fisher-Rao Information Metric in a natural way and show that, for the case of Gibbs equilibrium distributions, it coincides with the metric introduced in \cite{MNSS1990}.

\subsection{Differential entropy moments}

Let us consider a system whose macroscopic state is characterized by a set of  $n$  observables. Suppose that an experimentalist has measured the values of such observables up to some desirable accuracy. In such case the measurements are  identified with  the mean values
	\beq
	\label{meanvaluesgeneral}
	p_a \equiv \langle F_a \rangle  = \frac{\int_\Gamma  F_a \, \mu}{\int_\Gamma \mu} 
	\eeq
 of the set of  stochastic variables $\{F_a\}_{i=1}^n$.  Here $\Gamma$ is a sample space together with a well defined measure  $\mu$. In the case of a thermodynamic system $\Gamma$ is identified with  the phase space of statistical mechanics and $\mu$ is given in terms of an unassigned  probability distribution  $\rho:\Gamma\rightarrow \mathbb R^{+}$ such that
	\beq
	\label{PDF}
	{\rm vol}(\Gamma) = \int_\Gamma \mu = \int_\Gamma \rho\, \d x.
	\eeq
For instance, this is the case of the {  internal energy, which is} the average of the microscopic Hamiltonian. {  The choice of different controllable observables determines the particular statistical ensemble.} 

In the situation described above, the available information is given solely by the averages \eqref{meanvaluesgeneral} and the prescription for assigning the probability distribution  \eqref{PDF} is by means of the maximum entropy principle. Thus, let us introduce the \emph{microscopic entropy} of a generic distribution
	\beq
	\label{microentropy}
	s(\rho) =-\ln\rho\,.
	\eeq
whose weighted average yields {  (up to a normalization constant)} the entropy functional
	\beq
	\label{defentropy}
	S =\langle s(\rho) \rangle = -{\int_\Gamma \rho\, \text{ln}\rho\,\d x}\,.
	\eeq
		
%defined  over a sample space   $\Gamma$ , e.g. the statistical mechanics phase space of a given system. %\corrected{(for instance
%$\Gamma$ may be the phase space of statistical mechanics)} 
%and 
%	\beq
%	F^{i}:\Gamma\rightarrow \mathbb R^{+}, \quad i=1,\dots,n
%	\eeq
%a set of $n$ stochastic variables with respect to $\rho$. The Shannon entropy of a distribution $\rho$ is defined as 

The maximum entropy principle is expressed as a  constrained variational prescription for the functional
	\beq
	\tilde S =  \int_\Gamma \rho \ln \rho \, \d x - w\left( \int_\Gamma \rho\, \d x - 1\right) + q^a \left(\int_\Gamma F_a(x) \rho \, \d x  - p_a\right)\,,
	\eeq
where $w$ and $q^a$ are the corresponding Lagrange multipliers for the constraints given by
%As it is well-known from statistical mechanics, the distribution maximizing the entropy, subject to the constraints
	\begin{align}
	\label{eqcondition1}
	\int_\Gamma \rho\,\d x														  	& =	1\,,\\
	\label{eqcondition2}
	\frac{\int_\Gamma F_{a}(x)\,\rho\,\d x}{\int_\Gamma\rho\,\d x}	& =	p_{a} \qquad a=1,\dots,n \,.
	\end{align}
%\textcolor{red}{ADD EXAMPLE: $F^{1}=H$, $q^{1}=E$. Spend few words about ensembles.}

The result is the well known Gibbs distribution function 
	\beq
	\label{gibbsdistribution}
	\rho_{0}(\Gamma;w,q^{1},\dots,q^{n})=\text{e}^{-w+q^{a}F_{a}(x)}\,,
	\eeq
%where $w$ and $p_{1},\dots,p_{n}$ are the Lagrange multipliers corresponding to the constraints (\ref{eqcondition1}) and (\ref{eqcondition2}),
where  we have used Einstein summation convention.

One could then use the normalization condition (\ref{eqcondition1}) for the distribution $\rho_0$ (as it is usually the case in statistical mechanics), to set up the functional dependence
	\beq
	\label{gibbspartitionfunction}
	w(q^a)=\ln\int_\Gamma {\rm e}^{\,q^{a}F_{a}(x)}\d x = \ln \mathcal Z 
	\eeq
defining the partition function $\mathcal Z$ providing us with an  interpretation for $w$ as the \emph{free entropy} of the thermodynamic system and its derivatives with respect to the $q^{a}$ as the equations of state.
However, in what follows  we want to consider the full phase space of thermodynamics, so we need to have $w$ independent of the $q^{a}$, as we will make clear later.
Therefore, we will not consider the normalization condition -- equation \eqref{eqcondition1} -- but we will keep the form of $\rho_0$ to assign  our prior probability distribution so that  $w$ will be independent of the $q^{a}$.
 
%Let us define  the mean values of the stochastic functions with respect to a distribution $\rho$ as
%	\beq
%	\label{meanvaluesgeneral}
%	\langle F^{i}\rangle =\frac{\int F^{i}\,\rho\,\d\Gamma}{\int \rho\,\d\Gamma}\,.
%	\eeq
%The \emph{microscopic entropy} of a generic distribution $\rho$ is given by 
%	\beq
%	\label{microentropy}
%	s=-\text{ln}\rho\,.
%	\eeq
%\corrected{It follows from  (\ref{meanvaluesgeneral}) that } 
%	\beq
%	\label{microtomacrorhoG}
%	{\langle s(\rho)\rangle = \frac{\int s(\rho)\,\rho\,\d\Gamma}{\int \rho\,\d\Gamma}= - \frac{\int \rho \ln \rho\, \d \Gamma}{\int \rho\, \d \Gamma} = S(\rho)}\,,
%	\eeq
%where $S(\rho)$ is the macroscopic entropy of the distribution $\rho$ as defined in (\ref{defentropy}). Therefore, it is clear that the name \emph{microscopic entropy} for $s$ derives from the fact that its mean value is just the macroscopic entropy.  

Now we consider $\rho$ as a function on the $n+1$ control parameters $\lambda_{i}=w,q^{1},\dots,q^{n}$ and compute the differential of the microscopic entropy $\d s$, to obtain
	\beq
	\label{dsgeneral}
	\d s=-\frac{\partial {\rm ln}\rho}{\partial \lambda_{i}}\,\d \lambda_{i}.
	\eeq
Now, performing a moment expansion of the  differential entropy \eqref{dsgeneral}, it follows from  \eqref{meanvaluesgeneral} that  
	\beq
	\label{dsgeneralmeanvalue}
	\langle\d s\rangle=-\int_\Gamma  \left(\frac{\partial {\rm ln}\rho}{\partial \lambda_{i}}\,\d \lambda_{i}\right) \rho\,\d x =-\left\langle\frac{\partial {\rm ln}\rho}{\partial \lambda_{i}}\right\rangle \d\lambda_{i}
	\eeq
whilst the second moment yields
	\beq
	\label{dssquaredgeneralmeanvalue}
	\left\langle\left(\d s\right)^{2}\right\rangle =\int_\Gamma \left(\frac{\partial {\rm ln}\rho}{\partial \lambda_{i}}\, \frac{\partial {\rm ln}\rho}{\partial \lambda_{j}}\,\d \lambda_{i}\d \lambda_{j}\right) \rho \,\d x =
\left\langle\frac{\partial {\rm ln}\rho}{\partial \lambda_{i}}\frac{\partial {\rm ln}\rho}{\partial \lambda_{j}}\right\rangle \d \lambda_{i}\d \lambda_{j}\,.
	\eeq
	
Computing the averages in \eqref{dsgeneralmeanvalue} and \eqref{dssquaredgeneralmeanvalue} and using \eqref{gibbsdistribution} one obtains
	\beq
	\label{dsmeanvalue}
	\langle \d s \rangle = -\left\langle\frac{\partial \ln \rho}{\partial \lambda_i}\right\rangle \d \lambda_i  = \left\langle \d w  - F_a(x) \d q^a \right\rangle = \d w - p_a \d q^a.
	\eeq
%\textcolor{red}{THE SECOND EQUALITY ONLY HOLDS FOR GIBBS DISTRIBUTION \eqref{gibbsdistribution}.}
{  Using that the derivative of $\langle F_a\rangle$ with respect to the Lagrange multipliers $q^j$  [c.f. equation \eqref{eqcondition2}],
	\beq\label{diff_intensities1}
	\frac{\partial p_{a}}{\partial q^{b}}=\left\langle F_{a}\,F_{b}\right\rangle - p_{a}\,p_{b}=\left\langle \left(F_{a}-p_{a}\right) \left(F_{b}-p_{b}\right)\right\rangle\, ,
	\eeq
%is the covariance matrix  -- Cov($F_a,F_b$) -- 
it follows that  the variance of $\d s$ is }
	\beq
	\label{variances1}
	{\rm Var}(\d s)=\left\langle \left[\d s - \left\langle \d s\right\rangle\right]^{2}\right\rangle 
	= \left\langle \left(F_{a}-p_{a}\right) \left(F_{b}-p_{b}\right)\right\rangle\,\d q^{a}\,\d q^{b}\, 
	= \d p_{a}\,\d q^{a}\,,
	\eeq
where we have used equation (\ref{diff_intensities1}) to obtain the last identity, as it implies that 
	\beq
	\d p_{a}=\left\langle \left(F_{a}-p_{a}\right) \left(F_{b}-p_{b}\right)\right\rangle\,\d q^{b}.
	\eeq
Finally, using the well known formula for the variance
	\beq
	{\rm Var}(\d s) = \left\langle\left(\d s\right)^2\right\rangle - \left\langle \d s\right\rangle^2,
	\eeq
one obtains that the second moment of the microscopic entropy change is
	\beq
	\label{FisherRaorhoG1}
	\left\langle\left(\d s\right)^{2}\right\rangle = {\rm Var}(\d s) +\left\langle \d s\right\rangle^{2} = \d p_{a}\,\d q^{a}+\left(\d w-p_{a}\,\d q^{a}\right)^{2}.
	\eeq
	
Thus, the first moment of $\d s$ defines a 1-form field over an $n+1$ dimensional \emph{control manifold} $\mathcal{C}^{n+1}$ whose coordinates correspond to the control parameters $\lambda_i$. We will see in the next Section that promoting the $p_{a}$'s to independent variables, such a 1-form is an element of the co-tangent bundle of the Thermodynamic Phase Space introduced by Mruga{\l}a \cite{mrugala1} and that it defines its contact structure. Note that such a structure is obtained when using a generalized canonical equilibrium distribution of the form \eqref{gibbsdistribution}.	

The second moment \eqref{dssquaredgeneralmeanvalue} 
can be used to define a  metric tensor over the control manifold
\beq\label{FisherRao}
{G_{\rm FR}}=\sum_{i,j=1}^{n+1}{[G_{\rm FR}]}_{ij}\d \lambda_{i}\otimes\d \lambda_{j}=\left\langle\frac{\partial {\rm ln}\rho}{\partial \lambda_{i}}\frac{\partial {\rm ln}\rho}{\partial \lambda_{j}}\right\rangle\d \lambda_{i}\otimes\d \lambda_{j},
\eeq
explicitly given by
	\beq
	\label{FisherRaorhoG2}
	G_{\rm FR}(\rho)=\d q \otimes \d p+\left(\d w-p_{a}\,\d q^{a}\right)\otimes\left(\d w-p_{b}\,\d q^{b}\right)\,,
	\eeq
where 
	\beq
	\d q\otimes\d p=\frac{1}{2}(\d q^{a}\otimes\d p_{a}+\d p_{a} \otimes \d q^{a})\,,
	\eeq 
with $i,j=1,\dots,n$ and 
	\beq
	\d p_a = \frac{\partial p_a}{\partial q^b} \d q^b,
	\eeq
whose components are obtained from equation \eqref{diff_intensities1}.
 This result is well-known in the literature of statistical estimation theory, as it is the Fisher-Rao Information Metric \cite{BrodyRitz,Crooks}. Note that the position of the indices is conventional and we have adopted lower labels to distinguish the control parameters.

It is an important  remark that the Fisher-Rao metric \eqref{FisherRao} can be defined for \emph{any} distribution, i.e. independently if $\rho$ corresponds to the equilibrium distribution defined in 
\eqref{gibbsdistribution} or not (see \cite{Schlogl, BrodyRivier,BrodyRitz, Crooks}), and that equation \eqref{FisherRaorhoG2} gives its expression for a system in equilibrium, i.e. when $\rho=\rho_0$.
It is also worth emphasizing that in this Section we have considered the variables $p_{a}$ to be dependent solely on the Lagrange multipliers $q^{b}$, as it is clear from equations \eqref{diff_intensities1} and \eqref{variances1}. In the next Section we will assume that the $p_{i}$ are independent of the $q^{j}$ and write the metric \eqref{FisherRaorhoG2} in a $(2n+1)$-dimensional space
where all the variables $w,q^{a},p_{b}$ are independent, that is, in the Thermodynamic Phase Space.

% \corrected{watch out with the definition of $S$, to be changed? or is it the definition of the mean value that has to be changed? or everything can stay like this?}

\subsection{Relative Entropy}

Now, we review the meaning of relative entropy in the framework  we are pursuing here (see e.g. \cite{KLD}).
Let us fix the equilibrium distribution $\rho_{0}$ as the reference distribution and  compute the mean value of the microscopic entropy of
 another distribution $\rho\neq\rho_{0}$ with respect to $\rho_{0}$, 
that is,
	\beq
	\label{microtomacro}
	{\left\langle s(\rho)\right\rangle_{0} = \int_\Gamma s(\rho)\,\rho_{0}\,\d x = S(\rho_{0})-S(\rho,\rho_{0})}\,,
	\eeq
where $S(\rho,\rho_{0})$ is the relative entropy of $\rho$ with respect to $\rho_{0}$ (also called the Kullback-Leibler divergence of the two distributions \cite{KLD}), which 
is defined as 
	\beq
	\label{relativeentropydef}
	S(\rho,\rho_{0})=-\int_\Gamma \rho_{0}\,{\rm ln}\left({\rho_{0}}/{\rho}\right) \d x
	\eeq
and measures the loss of information that one gets when using the distribution $\rho$ instead of the proper one $\rho_{0}$ (see e.g. \cite{Schlogl}).
Therefore, we can obtain a characterization for the relative entropy of a distribution $\rho$ with respect to the equilibrium distribution $\rho_{0}$ as
	\beq
	\label{relentropy}
	S(\rho,\rho_{0})=\left\langle s(\rho_{0})\right\rangle_{0}-\left\langle s(\rho)\right\rangle_{0}=\left\langle \Delta s\right\rangle_{0}\,,
	\eeq
that is, the relative entropy provides us with  the mean difference between the microscopic entropies of the two distributions.
%\textcolor{red}{Say that for two different normalized exponential distributions this gives the difference in the free entropies, to make a link below.}

When the distribution $\rho$ is infinitesimally close to the equilibrium distribution $\rho_{0}$, we can write equation \eqref{relentropy}
in differential form. Up to first order terms we have
	\beq
	\label{relentropydiffform}
	S(\rho,\rho_{0}) = \left\langle s(\rho_{0})\right\rangle_{0}-\left\langle s(\rho)\right\rangle_{0}=\left\langle \d s(\rho_{0})\right\rangle_{0}\,.
	\eeq
However, as $S(\rho,\rho_{0})$ is always positive and has an absolute minimum for $\rho=\rho_{0}$ (with $S(\rho_{0},\rho_{0})=0$), 
expanding up to second order in the control parameters $\lambda_{i}$ yields
	\beq
	\label{relentropysecondorder}
	S(\rho,\rho_{0})=\frac{1}{2}\left\langle\frac{\partial^{2}s}{\partial \lambda_{i}\partial \lambda_{j}}\right\rangle_{0}\d \lambda_{i}\d\lambda_{j}=
\frac{1}{2}\left\langle\frac{\partial {\rm ln}\rho}{\partial \lambda_{i}}\frac{\partial {\rm ln}\rho}{\partial \lambda_{j}}\right\rangle_{0}\d \lambda_{i}\d \lambda_{j}\,.
	\eeq
Therefore, the components of Fisher-Rao metric are obtained as the Hessian of the function $S(\rho,\rho_{0})$ at the stationary point $\rho=\rho_{0}$. 
We will see in the next Section that the first order variation of $S(\rho,\rho_{0})$ defines a $1$-form while the second order variation yields a metric tensor for  the phase space of thermodynamics 
 endowing such manifold with a contact metric structure.

\section{The Contact and Riemannian structures of the Thermodynamic Phase Space}
\label{sec3}

Let us focus now only on the equilibrium distribution $\rho_0$. Motivated by the first moment of the relative entropy [c.f. equation \eqref{dsmeanvalue}],
 we promote the mean values $p_i$  defined in \eqref{eqcondition2} as independent coordinates 
 of a larger manifold which we call the Thermodynamic Phase Space and denote it  by $\mathcal{T}^{2n +1}$. 
 Such a space has naturally  $2n+1$ dimensions, $n+1$ corresponding to the control variables and $n$ to the  
 mean values $p_i$ corresponding to each Lagrange multiplier $q^i$. In this manner, equation \eqref{dsmeanvalue} 
 becomes a contact 1-form for the Thermodynamic Phase Space  $\mathcal T^{2n+1}$ in a Darboux patch whose local coordinates are given  by $w,q^{1},\dots,q^{n},p_{1},\dots,p_{n}$. 

Let us clarify our opening statement through a revision of some geometric theories of thermodynamics \cite{mrugala1,mrugala2,GTD,CRGTD,HernLaco}.  
Just as in its symplectic version, Darboux's theorem states that around each point of a contact manifold, 
there is a coordinate patch in which the  1-form defining the contact structure of $\mathcal{T}^{2n+1}$ reduces to 
\beq\label{eta}
\eta\equiv\d w-p_{i}\,\d q^{i}, \quad \text{with} \quad \mathcal{D} = \ker(\eta), 
\eeq
where $\mathcal{D}$ is the contact distribution of $\mathcal{T}^{2n+1}$ \cite{Arnold}. Since the coordinates $w,q^{1},\dots,q^{n},p_{1},\dots,p_{n}$ are all independent in the Thermodynamic Phase Space, the $1$-form  \eqref{eta} is non-degenerate and satisfies the maximal non-integrability condition
\beq
{\rm vol}_\eta \equiv \eta\wedge(\d\eta)^{n}\neq0.
\eeq
Therefore it is a well defined  volume form on $\mathcal T^{2n+1}$.

The spaces of thermodynamic equilibrium states corresponding to particular systems 
are identified with the Legendre sub-manifolds of the contact distribution $\mathcal{D}$ defined by $\eta$. 
That is, the $n$ dimensional sub-manifolds $\mathcal{E}^n$ embedded in $\mathcal{T}^{2n+1}$ whose tangent bundle is completely contained in $\mathcal{D}$. It is easy to see that if
	\beq
	\varphi: \mathcal{E}^n \longrightarrow \mathcal{T}^{2n+1}
	\eeq
is one such embedding, then $\mathcal{E}^n$  is locally defined through the equation 
\beq\label{eta_equals_zero}
\varphi^*\eta=\varphi^*[\d w-p_{i}\,\d q^{i}]=0\,,
\eeq
where 
	\beq
	\label{pullbackE}
	\varphi^*: T^* \mathcal{T}^{2n+1} \longrightarrow T^*\mathcal{E}^n
	\eeq
is the induced map associated with $\varphi$ and \eqref{eta_equals_zero} provides us with the (local) explicit form of the embedding $\varphi:[p_i] \rightarrow [w(q^i), p_{j}(q^i), q^i]$, that is
	\beq
	\label{eos1}
	p_j(q^i) = \frac{\partial w}{\partial q^j},
	\eeq
where the only freedom rests upon the choice of $w=w(q^i)$. 

One can readily see that \eqref{eta_equals_zero} is a local statement of the first law for a system described by the fundamental relation $w$. 
This construction has been used as the basis for various programmes in geometric thermodynamics, albeit most of these programmes have focused only on the choice of a metric tensor for the Legendre sub-manifold $\mathcal{E}^n$ \cite{wein1975,rupp1979,rupp1995}. 
In the construction presented here, the metric tensor for the equilibrium space $\mathcal{E}^n$ will be obtained by means of the induced map \eqref{pullbackE} of the metric tensor for $\mathcal{T}^{2n+1}$ 
	\beq
	\label{metricT}
	G = \eta \otimes \eta +\frac{1}{2} \left( \d q^i \otimes \d p_i + \d p_i \otimes \d q^i \right).
	\eeq
Note that this choice is not arbitrary. It is constructed so that the pull-back of an embedding $\phi$ of  the control manifold $\mathcal{C}^{n+1}$ into $\mathcal{T}^{2n+1}$
	\beq
	\label{embebddingCT}
	\phi: (w,q^i) \longrightarrow (w, p_j(q^i), q^i).
	\eeq
coincides with  the Fisher-Rao metric on $\mathcal{C}^{n+1}$ [c.f. equations \eqref{FisherRao}-\eqref{FisherRaorhoG2}], that is
	\beq
	G_{\rm FR} = \phi^*(G).
	\eeq
Moreover, one can directly verify that
	\beq
	\phi^*(\eta) = \langle \d s \rangle_{0}.
	\eeq
 Thus, at every point of the control manifold, the 1-form field $\langle \d s \rangle_{0} \in T^*\mathcal{C}^{n+1}$ is annihilated by vectors lying on the contact distribution $\mathcal{D}$ of $\mathcal{T}^{2n+1}$. In this sense, the Legendre sub-manifolds of the Thermodynamic Phase Space correspond to equilibrium states maximizing the averaged microscopic entropy as long as the map
	\beq
	\label{pullfirst}
	(\phi^{-1} \circ \varphi)^* \langle \d s\rangle_{0} = 0
	\eeq
is well defined. Locally this is always the case provided the embedding  $\phi:\mathcal{C}^{n+1} \longrightarrow\mathcal{T}^{2n+1}$ is a $C^1$ invertible map {  at any point $\mathcal{P}$}. 
That is, if the linear term in the expansion
	\beq
	\label{locexp}
	p_j(q^i) = p_j(\mathcal{P}) + \Delta q^i\left(\frac{\partial p_{j}}{\partial q^i}\right)_{\mathcal{P}} + \mathcal{O}\left[(\Delta q^i)^2\right]
	\eeq
is non-vanishing.  Thus, equation \eqref{pullfirst} defines an equivalence, to linear order, between Legendre sub-manifolds and stationary points of the averaged microscopic entropy. 
In this sense, equation \eqref{pullfirst} is a re-statement of the First Law of thermodynamics. The situation can be summarized by
	\beq
	\includegraphics[width=.4\columnwidth]{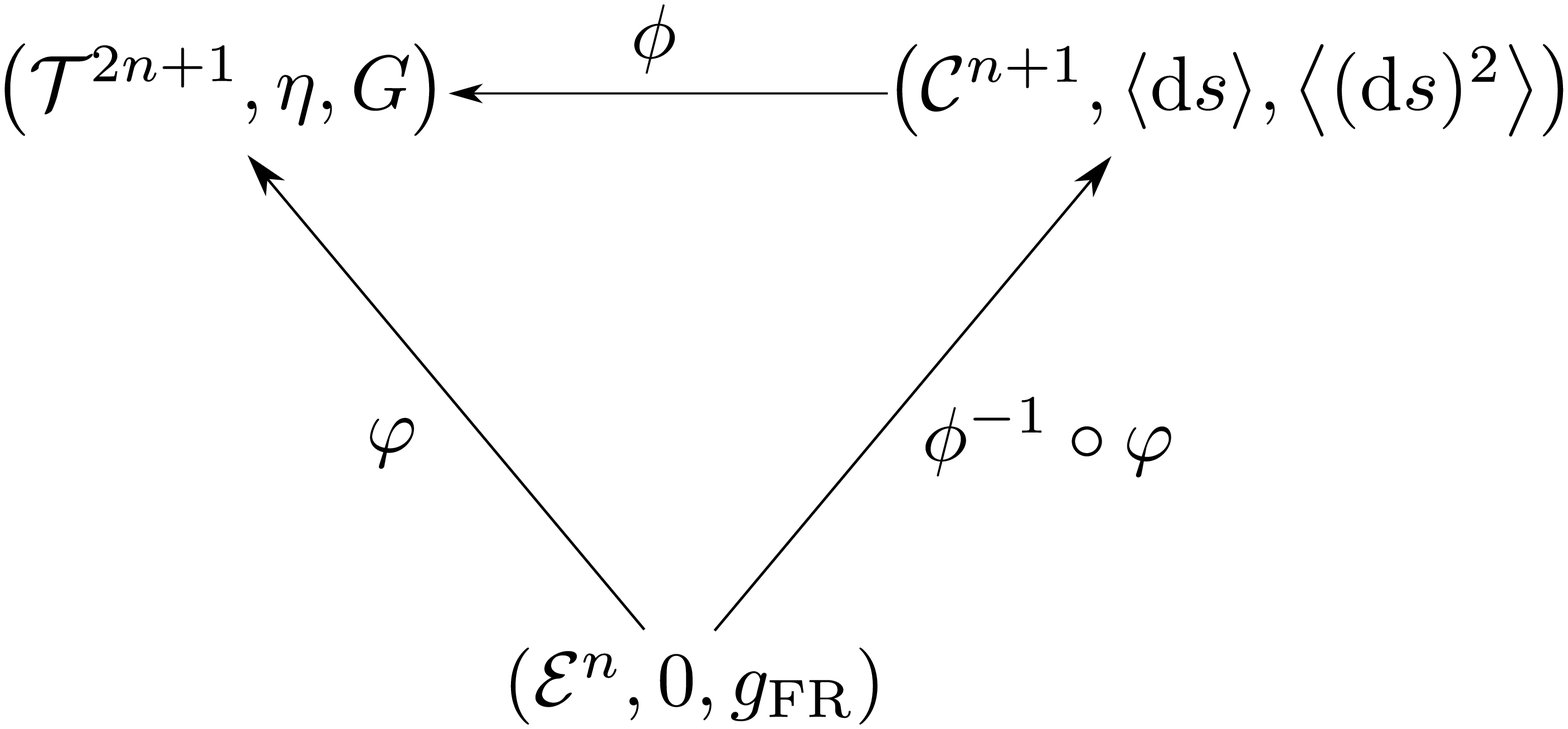}	
	\eeq

The local invertibility of $\phi$ can be interpreted as the existence of local equilibrium, independently of  any particular system characterized by a fundamental relation $w(q^i)$, defining the embedding $\varphi$. Moreover,  taken all the way down to $\mathcal{E}^n$, the condition of local invertibility  of $\phi$, equation \eqref{locexp}, becomes the condition
	\beq
	\det\left(\frac{\partial^2 w}{\partial q^i \partial q^j}\right) \neq 0,
	\eeq
which is equivalent to demanding that the induced Fisher-Rao metric on $\mathcal{E}^n$
	\beq
	\label{indFR}
	g_{\rm FR} = \varphi^*(G) = \frac{\partial^2 w}{\partial q^i \partial q^j} \, \d q^i \otimes \d q^j\, ,
	\eeq
is non-degenerate.

%and phase transitions are regions where different leaves of this foliation intersect each other \cite{Arnold, Legendresymmetry}.
It is worth noticing here that we have obtained the universality of local equilibrium  as a direct  consequence of the extremization of 
the entropy functional ($\rho_0$ is the distribution maximizing the entropy) or, alternatively, as the condition 
of the vanishing of the first order variation of the relative entropy $S(\rho,\rho_0)$, see Eq. \eqref{relentropydiffform}.
%Thus,  Eq. \eqref{eta_equals_zero}   microscopic statement of the first law of thermodynamics, that is, during an equilibrium (quasi-static)
%process the entropy of the distribution does not change, up to first order. [REPHRASE] 
%In fact, since the macroscopic entropy $S$ is extremized by $\rho_0$, it follows from 
%(\ref{dsmeanvalue})
%that 
%$\eta=0$ if and only if  $\langle \d s\rangle =0$, Eq. \eqref{eta_equals_zero2}.
Moreover, the components for the Fisher-Rao metric correspond to the second moment $\langle(\d s)^{2}\rangle_{0}$, c.f. equation \eqref{FisherRaorhoG2}. At equilibrium, 
the Second Law implies that the entropy must be a maximum. Therefore, when $w$ is identified as  the entropy, the induced Fisher-Rao metric, equation \eqref{indFR}, has to be negative definite.
Notice that in such case the metric \eqref{indFR} coincides with the thermodynamic metric introduced by Ruppeiner in the context of thermodynamic fluctuation theory \cite{rupp1995,rupp2010,rupp2012} up to a sign.

%where $w(p_{i})$ is the free entropy $w\equiv-\beta F={\rm ln}Z$. In this manner, the components of the metric $g$ are given by the Hessian of the free entropy. Therefore, the metric $g$ is also equivalent to Ruppeiner's thermodynamic metric \cite{MNSS1990}. In fact, as it was shown in \cite{Schlogl}, the two metrics are different over finite displacements but they agree over infinitesimal ones and they are related in general by a Legendre Transform \cite{Schlogl,BrodyRivier,Crooks}. Moreover, the crucial point in our discussion is that the metric \eqref{FisherRaorhoG2} can be defined for \emph{any} distribution $\rho$
%[see Eq. \eqref{FisherRao}] and, as such, is best qualified to account for further use in general statistics and in the study of thermodynamic systems out of equilibrium \cite{Schlogl,BrodyRivier,Crooks,CrooksPRE2012,CrooksPRL2012}.
%Finally, $G$ gives a geometric meaning to the second law of thermodynamics. In fact, the second law of thermodynamics implies that $g={G}|_{\mathcal E}$ must be positive definite.

We stress here the fact that this construction on the Thermodynamic Phase Space is completely general and holds for \emph{any} thermodynamic system. This means 
that the forms of $\eta$ and $G$ do not change from system to system, as well as the First and Second Law of thermodynamics apply in full  generality. The specification of a particular equilibrium system 
  corresponds to a particular choice of the fundamental function $w(p_{i})$. The First Law and the equations of state  follow from Eq. (\ref{eta_equals_zero}). In this way, the induced metric 
$g_{\rm FR}$  on each Legendre sub-manifold  $\mathcal{E}^n$ is specified for each particular system.

%Let us close this section commenting on Mrugala et al. \cite{MNSS1990}. Here, we are presented with some open problems related to the group of symmetries of the contact Riemannian manifold $(\mathcal T^{2n+1},\eta,{G})$ that are relevant for the geometric theory of thermodynamics, the interpretation of the metric $G$ on the Thermodynamic Phase Space and the possible applications of this formalism to understand the geometry of non-equilibrium phenomena.  Although there has been some attempts to understand such points,  the full answer to the these issues on the geometric nature of the TPS is still lacking. In the next Section we will give an answer to some of these questions. In particular, we will show that the contact Riemannian manifold $(\mathcal T^{2n+1},\eta, G)$ is, in fact, a Sasakian manifold and that its local structure group (reduced from the group of contactomorphisms) is $U(n) \times 1$ (c.f. reference \cite{bookBlair}). Therefore  $U(n)\times 1$ is the relevant group of symmetries for the local theory of thermodynamics, akin to the Lorentz group for the General Theory of Relativity. Moreover, we will also show that $(\mathcal T^{2n+1},\eta,G)$ is an $\eta$-Einstein manifold. This last observation will be relevant for discussions about extensions to non-equilibrium systems.

%\begin{figure}
%\begin{center}

%\caption{Diagram}
%\end{center}
%\end{figure}

\section{Geometric properties of the phase space of Equilibrium Thermodynamics}\label{sec4}

In this Section we study the geometry of the Thermodynamics Phase Space in greater detail. 
As we have previously discussed, such a space is a metric contact manifold whose contact and metric structures 
are defined through the mean value and the second moment  of the infinitesimal  change in the microscopic entropy weighted by the equilibrium Gibbs distribution, 
respectively [c.f. equations \eqref{dsmeanvalue} and \eqref{FisherRaorhoG1}].  

 In general, a contact 1-form $\eta$ is not unique, but it  belongs to the class generating the 
 contact structure $\mathcal{D}$ of $\mathcal{T}^{2n+1}$, c.f. equation \eqref{eta}. Indeed, 
 any other 1-form defining the same ${\mathcal D}$  is necessarily  conformally equivalent to $\eta$,
 i.e. for any two 1-forms $\eta_1$ and $\eta_2$ in the same equivalence class $[\eta]$, one has
$ \eta_2 = \Omega \eta_1$
for some non-vanishing real function $\Omega$.

 For each member in the class generating $\mathcal{D}$  there is a unique canonical vector field $\xi$, called the \emph{Reeb vector field},  satisfying 
\beq\label{ReebVF}
\eta(\xi)=1 \quad \text{and} \quad \d \eta(\xi,X)=0\,,
\eeq
for any vector field $X\in T\mathcal{T}^{2n+1}$. 
The Reeb vector field generates a splitting of the tangent bundle $T \mathcal{T}^{2n+1}$, that is 
\beq\label{splitting}
T\mathcal{T}^{2n+1} = L_{\xi}\oplus \cal D\,,
\eeq
where $ L_{\xi}$ is the sub-space generated by $\xi$. 

 It will be convenient to work in a basis adapted to the splitting \eqref{splitting}. Since we have already seen that the Reeb vector field generates the vertical part, it only remains to find a basis for the contact distribution. Choosing Darboux local coordinates, it is easy to see that 
	\beq
	\eta\left(\hat Q^b \right) \equiv \left[ \d w - p_a \d q^a \right]\left( \frac{\partial }{\partial q_b}+p_b \frac{\partial }{\partial w}  \right) = p_b - p_a \delta^{a}_{\ b} = 0
	\eeq
and, similarly,
	\beq
	\eta\left(\hat P_b\right) \equiv  \left[ \d w - p_a \d q^a \right] \left(\frac{\partial }{\partial p_b} \right) = 0 . 
	\eeq
Therefore, the vectors
	\beq\label{Dbasis}
	 \left\{\frac{\partial}{\partial p_a}, \frac{\partial}{\partial q^a} +p_a \frac{\partial}{\partial w} \right\} \subset \ker{(\eta)}
	\eeq
are linearly independent and generate $\mathcal{D}$. Thus, the non-coordinate basis
	\beq
	\label{basisTT}
	\Big{\{}\xi,\hat{Q}_{a},\hat{P}^{a}\Big{\}} = \left\{\xi\,, \frac{\partial}{\partial q^a}+p_a\frac{\partial }{\partial w}\,, \frac{\partial }{\partial p_a} \right\}
	\eeq
 is naturally adapted to the splitting \eqref{splitting} induced by the gauge choice $\eta \in [\eta]$. Notably,  the generators of such  basis satisfy the commutation relations
	\beq
	\label{halgebra}
	[\hat P^a,\hat Q_b] = \delta^a_{\ b} \xi, \quad [\xi,\hat Q_a] = 0 \quad \text{and} \quad [\xi, \hat P^a] = 0,
	\eeq
with respect to the Lie-bracket and where $\delta^{a}_{\ b}$ represents the $n\times n$-Kronecker delta. 
These are the defining relations of the  $n$th Heisenberg Lie algebra $\mathcal{H}_n$. 
For this reason, we call the set  \eqref{basisTT} \emph{the Heisenberg basis} of $T\mathcal{T}$. 
Note that the above commutation relations arose naturally from the definition of the contact structure $\mathcal{D}$ motivated by the mean value of the microscopic entropy change.

Now,  we will show that by taking into account the macroscopic information stemming from the second moment of $\d s$ -- equation \eqref{FisherRao}  -- 
the Thermodynamic Phase Space is uniquely defined as the hyperbolic Heisenberg group defined in \cite{IVZ}. 
That is, we will show that the Thermodynamic Phase Space is a para-Sasakian manifold with a flat canonical connection. 
To this end, we verify that the metric \eqref{metricT}  satisfies  some formal definitions following the construction in \cite{Zamkovoy} and \cite{Perrone}.

We have already selected a 1-form in the class defining the contact structure of the Thermodynamic Phase Space 
and equipped this manifold with a metric structure given by \eqref{metricT}. That is, we have the quadruple $(\mathcal{T},\eta,\xi,G)$. If, in addition, there is a $(1,1)$-tensor field $\Phi$ 
satisfying
	\beq
	\label{ACS_prop}
	L_\xi = {\rm ker}(\Phi), \quad  \mathcal{D} = {\rm Im}(\Phi) \quad \text{and} \quad \Phi^2 = {\rm I} - \eta \otimes \xi,
	\eeq
such that 
	\beq\label{almostparadef}
	\eta(X) = G(\xi, X), \quad \text{and} \quad \frac{1}{2}\d \eta(X,Y) = -G(X, \Phi Y) 
	\eeq
 for any pair of vector fields $X$ and $Y$, we call $\Phi$ an \emph{almost-para-contact structure} and $(\mathcal{T},\eta,\xi,\Phi,G)$ a \emph{para-contact metric manifold} 
 \footnote{Usually in the literature there is a slightly different definition for a para-contact metric manifold (see e.g. \cite{Zamkovoy,Perrone}), due to the fact that the authors take 
 $\d\eta(X,Y)=\frac{1}{2}\left(X\eta(Y)-Y\eta(X)-\eta([X,Y])\right)$,
 which in fact differs in our case from the exterior derivative of $\eta$ by a factor $1/2$. Taking into account this difference in $\d\eta$, one finds that the two definitions coincide.}.
 
 To determine the form of $\Phi$ let us consider the Levi-Civita connection associated with $G$.  Then, the covariant derivative of the Reeb vector field satisfies \cite{Zamkovoy}
	\beq
	\label{apc}
	\nabla \xi = - \Phi + \Phi h
	\eeq
where
	\beq
	h = \frac{1}{2}\, \mathcal{L}_\xi \Phi.
	\eeq
If $\xi$ is a Killing vector {of $G$}, the tensor field $h$ vanishes identically. Thus,  equation \eqref{apc} directly defines the almost-para-contact structure. 
Working in Darboux coordinates, the Reeb vector field is simply given by
%	\beq
$	\xi = \frac{\partial }{\partial w}.$
%	\eeq
Thus, since none of the metric components in \eqref{metricT}
 is a function of $w$, $\xi$ is indeed a Killing vector of $G$ and $\Phi = -\nabla \xi$.   
 Moreover, the vectors in the Heisenberg basis  generating the horizontal distribution are all  null with respect to the metric $G$, that is
	\beq
	G(\hat P^a,\hat P^a) = G(\hat Q_a, \hat Q_a) = 0.
	\eeq
Thus, the metric has $n+1$ `space-like' directions and $n$ `time-like' directions, i.e. the signature of $G$ is $(n+1,n)$.  
To make explicit  the pseudo-Riemannian  signature of the metric,  let us introduce  the orthonormal (dual) basis
	\beq
	\label{Obasis}
	\hat \theta^{(i)} = \left\{\hat \theta^{(0)}, \hat \theta^{(a)}_{+},\hat \theta^{(a)}_{-} \right\} \quad i =0\dots 2n,
	\eeq
where
	\beq
	\hat \theta^{(0)} = \eta \quad \text{and} \quad \hat  \theta^{(a)}_{\pm} = \frac{\sqrt{p_a}}{2 p_a}\left[  p_a \d q^a  \pm  \d p_a\right] \quad \text{(no sum over $a$)}.
	\eeq
In this case, the metric \eqref{metricT} can be written as
	\beq
	\label{metricC}
	G = \hat \theta^{(0)} \otimes \hat \theta^{(0)} + \sum_{a=1}^n \left[\hat \theta^{(a)}_{+} \otimes \hat\theta^{(a)}_{+} - \hat\theta^{(a)}_{-} \otimes \hat\theta^{(a)}_{-}\right],
	\eeq
whose matrix representation is 
	\beq
	G_{ij} = \left(
							\begin{array}{ccc}
							1 & 0 &  0 			\\
							0 & \delta^{a}_{\ b} & 0 \\
							0 & 0 & -\delta^{a}_{\ b} \\
							\end{array}
							\right)\,.
	\eeq
 Thus, in this convention, the $n$ `time-like' directions are given by 
		\beq
		\hat e_{(a)}^- = G^{-1}\left[\hat \theta^{(a)}_{-}\right] 
		%= \sqrt{p_a} \left(\xi - \frac{\partial}{\partial p_a} \right) + \frac{1}{\sqrt{p_a}} \frac{\partial}{\partial q^a} 
		={\sqrt{p_{a}}}\left[\frac{1}{p_{a}}\hat Q_{a}-\hat P^{a}\right]
		\quad \text{(no sum over $a$)},
		\eeq
while the $n+1$ `space-like' directions are
		\beq
		\hat e_{(0)} = \xi \quad \text{and} \quad \hat e_{(a)}^+ 
		= G^{-1} \left[\hat \theta^{(a)}_{+}\right] 
		%= \sqrt{p_a} \left(\xi + \frac{\partial}{\partial p_a} \right) + \frac{1}{\sqrt{p_a}} \frac{\partial}{\partial q^a} 
		={\sqrt{p_{a}}}\left[\frac{1}{p_{a}}\hat Q_{a}+\hat P^{a}\right]
		\quad \text{(no sum over $a$)}.
		\eeq
Therefore,  let us define the non-coordinate basis
		\beq
		\label{canbasis}
		\hat e_{(i)} = \left\{\hat e_{(0)}, \hat e_{(a)}^+,\hat e_{(a)}^- \right\} \qquad (i=0\dots 2n)
		\eeq
whose structure functions can be read from the only non-vanishing Lie-brackets
	\beq
	\label{strucF}
	\left[\hat e^+_{(a)}, \hat e^-_{(a)} \right] = - \frac{1}{2 \sqrt{p_a}} \left(\hat e^+_{(a)} + \hat e^-_{(a)} \right) + 2 \hat e_{(0)}  \qquad (a= 1\dots n).
	\eeq
We call \eqref{canbasis} the \emph{canonical basis of the Thermodynamic Phase Space}. 
Throughout the rest of the paper, all the calculations will be performed with respect to this basis. Note that in our convention, the indices  $i,j,k$ vary from $0$ to $2n$ while $a,b,c$ take values from $1$ to $n$.  

First, let us note that the non-vanishing the Levi-Civita connection symbols in this  non-coordinate basis are
	\beq
	\label{Chris1}
	\Gamma^0_{(n+a) \ (a)} = -\Gamma^0_{a \ (n+a)} =  \Gamma^a_{0 \ (n+a)} = \Gamma^a_{(n+a) \ 0} = \Gamma^{n+a}_{0 \ a} = \Gamma^{n+a}_{a \ 0} = 1 ,
	\eeq
	\beq
	\label{Chris2}
	\Gamma^a_{(n+a)\ (n+a)} = \Gamma^{n+a}_{a\ (n+a)}  = -\Gamma^{a}_{(n+a) a} = -\Gamma^{n+a}_{a \ a} = \frac{1}{2\sqrt{p_a}}.
	\eeq

Now, an expression for the almost-para-contact structure is direclty obtained from \eqref{apc} whose form in the canonical basis is
	\begin{align}
	\label{acsC}
	\Phi = - \nabla \hat e_{(0)} 	& =- \sum_{i,j=0}^{2n}\Gamma^i_{j0} \ \hat e_{(i)} \otimes \hat \theta^{(j)}\nonumber\\
													& = -\sum_{j=0}^{2n} \sum_{a=1}^n\left[\Gamma^a_{j0} \ \hat e^+_{(a)} \otimes \hat \theta^{(j)}+ \Gamma^{n+a}_{j 0} \ \hat e^-_{(a)} \otimes \hat \theta^{(j)}\right]\nonumber\\
													& = -\sum_{a=1}^n\left[\Gamma^a_{(n+a) \ 0} \ \hat e^+_{(a)} \otimes \hat \theta^{(a)}_- + \Gamma^{n+a}_{a \ 0} \ \hat e^-_{(a)} \otimes \hat \theta^{(a)}_+\right]\nonumber\\
													& = -\sum_{a=1}^n \left[\hat e_{(a)}^+ \otimes \hat\theta^{(a)}_{-} + \hat e_{(a)}^{-} \otimes \hat\theta^{(a)}_{+} \right].
	\end{align}
Indeed,  $\Phi^2$  satisfies
	\begin{align}
		\Phi^2		& =  -\left[\hat e_{(a)}^+ \otimes \hat\theta^{(a)}_{-} + \hat e_{(a)}^{-} \otimes \hat\theta^{(a)}_{+} \right] \left(  -\left[\hat e_{(b)}^+ \otimes \hat\theta^{(b)}_{-} + \hat e_{(b)}^{-} \otimes \hat\theta^{(b)}_{+} \right]\right)\nonumber\\
						& =\left[ \hat e^+_{(a)} \otimes \hat \theta^{(a)}_{-} \right] \left(\hat e_{(b)}^+ \otimes \hat\theta^{(b)}_{-} + \hat e_{(b)}^{-} \otimes \hat\theta^{(b)}_{+} \right) +\left[\hat e^-_{(a)} \otimes \hat \theta^{(a)}_{+} \right]\left(\hat e_{(b)}^+ \otimes \hat\theta^{(b)}_{-} + \hat e_{(b)}^{-} \otimes \hat\theta^{(b)}_{+} \right) \nonumber\\
						& =\left[ \delta^a_{\ b} \hat e^+_{(a)} \otimes \hat \theta^{(b)}_{+}\right] + \left[\delta^a_{\ b} \hat e^-_{(a)} \otimes \hat \theta^{(b)}_{-}\right]\nonumber\\
						& = \left[\hat e^+_{(a)} \otimes \hat \theta^{(a)}_{+}\right] + \left[ \hat e^-_{(a)} \otimes \hat \theta^{(a)}_{-}\right] 
	\end{align}
where we have used the Einstein sum convention. Its matrix expression is
	\beq
	\left[\Phi^2\right]^i_{\ j} =\left[\begin{array}{ccc}
													0 & \cdots & 0 \\
													0 & \delta^a_{\ b} & 0\\
													0 & \cdots & \delta^a_{\ b}
													\end{array}
													\right]
	\eeq
 and, from \eqref{acsC}, the action of $\Phi$ on the elements of the basis is trivially given by
	\begin{align}
	\label{apc_action1}
	\Phi  \hat e_{0} & = 0,\\
	\Phi \hat e_{(a)}^+ &= -\hat e_{(a)}^-,\\
	\label{apc_action2}
	\Phi \hat e_{(a)}^- &= - \hat e_{(a)}^+\,.
	\end{align}
Thus, the defining requirements of \eqref{ACS_prop} are fulfiled.
 
The metric $G$ is -- by construction -- compatible and associated to the almost-para-contact structure $\Phi$. 
Let us revise these concepts explicitly by means of a pair of arbitrary vector fields 
	\beq
	\label{XY1}
	X= X^i \hat e_{(i)} = X^0 \hat e_{(0)} + \sum_{a=1}^n \left[X^a_+ \hat e^+_{(a)}  + X^a_- \hat e^-_{(a)}\right].
	\eeq
and
	\beq
	\label{XY2}
	Y= Y^i \hat e_{(i)} = Y^0 \hat e_{(0)} + \sum_{a=1}^n \left[Y^a_+ \hat e^+_{(a)}  + Y^a_- \hat e^-_{(a)}\right].
	\eeq
It follows from \eqref{apc_action1} - \eqref{apc_action2} that 
	\beq
	\label{apcX}
	\Phi X = - \sum_{a=1}^n\left[X^{a}_- \hat e_{(a)}^+ + X^{a}_+ \hat e_{(a)}^-\right],
	\eeq
	\beq
	\label{apcY}
	\Phi Y = - \sum_{a=1}^n\left[Y^{a}_- \hat e_{(a)}^+ + Y^{a}_+ \hat e_{(a)}^-\right].
	\eeq
The inner product of the vector fields \eqref{apcX} and \eqref{apcY} induced by the metric \eqref{metricC} is given by
	\beq
	\label{compa1}
	G(\Phi X, \Phi  Y) = \sum_{a=1}^n \left[ X^{a}_- Y^{a}_- - X^a_+ Y^a_+ \right],
	\eeq
while that of \eqref{XY1} and \eqref{XY2} is
	\beq
	\label{compa2}
	G(X,Y) =  X^0 Y^0 - \sum_{a=1}^n \left[ X^{a}_- Y^{a}_- - X^a_+ Y^a_+ \right].
	\eeq
We say that the metric $G$ is \emph{compatible with the almost para-contact structure $\Phi$} if the condition 
	\beq
	G(\Phi X, \Phi Y) =  -G(X,Y) + \eta(X) \eta(Y)
	\eeq
is satisfied. Thus, it follows from \eqref{compa1} and \eqref{compa2}, 
together with the obvious fact that $\eta(X) = X^0$ and $\eta(Y) = Y^0$, that $G$ -- given by \eqref{metricC} -- 
is compatible with $\Phi$ -- computed in \eqref{acsC}. Moreover, the metric $G$ is an {associated metric to the almost-para-contact structure}, that is
	\beq
	\label{asocC0}
	G(X,\Phi Y) = -\frac{1}{2} \d \eta (X,Y)
	\eeq
is satisfied -- c.f. definition \eqref{almostparadef}. In the canonical basis, the exterior derivative of the contact 1-form takes the form
	\beq
	\d \eta = -2 \sum_{a=1}^n \left[\hat \theta^{(a)}_+ \wedge \hat\theta^{(a)}_- \right].
	\eeq
Thus, indeed, 
	\beq
	\label{asocC}
	G(X,\Phi Y ) = - \sum_{a=1}^n \left[ X^{a}_+ Y^{a}_- - X^{a}_- Y^{a}_+ \right] = - \frac{1}{2} \d \eta(X,Y).
	\eeq
A contact manifold endowed with a Riemannian structure such that \eqref{asocC0} is satisfied is called an \emph{para-contact metric manifold}. 
Thus, the Thermodynamic Phase Space $(\mathcal{T}^{2n+1},\eta,\xi,\Phi,G)$ is a para-contact metric manifold.

Next, we verify that the Thermodynamic Phase Space is, in fact, a \emph{para-Sasakian} manifold. 
To this end, we need to show that the structure $(\mathcal{T}^{2n+1},\eta,\xi,\Phi,G)$ is \emph{normal}, i.e. that the Nijenhuis tensor of the almost-para-contact structure 
	\beq
	N_\Phi (X,Y)  \equiv \Phi^2 [X,Y] + [\Phi X,\Phi Y] - \Phi [\Phi X,Y] - \Phi [X,\Phi Y]
	\eeq
satisfies the condition
	\beq
	\left[N_\Phi +  \d \eta \otimes \xi\right] (X,Y) = 0 .
	\eeq
A long but straightforward calculation yields
	\beq
	\label{Nijenhuis}
	N_\Phi (X,Y) = -2 \sum_{a=1}^n \left[X^a_+ Y^{a}_- -X^{a}_- Y^a_+ \right] \hat e_{(0)}
	\eeq
and from the second equality of \eqref{asocC}, it follows that the normality condition is satisfied.  This is a statement of the integrability of the almost-para-complex structure of the horizontal distribution, that is, the restriction of the Nijenhuis tensor to the horizontal space $\mathcal{D}$ vanishes identically, as can be seen from \eqref{Nijenhuis}. In this case, we say that the structure $(\mathcal{T}^{2n+1},\eta,\xi,\Phi,G)$ is \emph{integrable}.

Finally, we use the result of Ivanov, Vassilev and Zamkovoy (IVZ)  which states that 
an integrable para-contact metric manifold of dimension $2n+1$ is locally isomorphic to the hyperbolic Heisenberg group 
exactly when the canonical connection $\tilde\nabla$ has vanishing {  horizontal} curvature, {  i.e. $\tilde R(X,Y,Z,V)=0$ for all $X,Y,Z,V \in \mathcal{D}$}  {  (c.f. IVZ Theorem in Appendix \ref{appB})}. 
Here, the \emph{canonical connection} refers to the one compatible with all the objects defining the para-Sasakian structure, namely, the one satisfying
	\beq
	\tilde\nabla \eta = \tilde\nabla \xi = \tilde\nabla \Phi = \tilde \nabla G = 0.
	\eeq
and whose torsion satisfies
	\beq
	\tilde T (\xi, \Phi Y) = - \Phi \tilde T(\xi,Y)
	\eeq
and
	\beq
	\tilde T(X,Y) = 2 \d \eta(X,Y) \xi.
	\eeq
%	\textcolor{red}{MISSING SECOND CONDITION ON THE TORSION}

On an integrable para-contact metric manifold such a connection is unique and is defined in terms of the Levi-Civita connection by (c.f. Equation (4.44) in \cite{Zamkovoy})
	\beq\label{canonicalconn}
	\tilde \nabla_X Y = \nabla_X Y + \eta(X) \Phi Y - \eta(Y) \nabla_X \xi + \left(\nabla_X \eta \right)( Y ) \xi.
	\eeq
Working out the connection symbols of $\tilde\nabla$ with respect to the canonical basis,
	\beq
	\tilde \nabla _{\hat e_{(i)}} \hat e_{(j)} = \tilde\Gamma^{k}_{ji} \hat e_{(k)} \qquad (i,j,k=0\dots 2n),
	\eeq
it follows that
	\beq
	\label{TW1}
	\tilde \Gamma^0_{ji} = - \delta^0_{\ i} \Gamma^0_{0j} - \delta^0_{\ j} \Gamma^0_{0 i} 	= 0
	\eeq
and
	\beq
	\label{TW2}
	\tilde \Gamma^\alpha_{ji}  = \Gamma^\alpha_{ji} -   \delta^0_{\ i} \Gamma^\alpha_{0j} - \delta^0_{\ j} \Gamma^\alpha_{0 i}  \qquad (\alpha=1\dots 2n),
	\eeq
where Levi-Civita connection symbols are given by \eqref{Chris1} and \eqref{Chris2}. Thus, the only non-vanishing connections symbols are
	\beq
	\label{nonzeroC}
	 - \tilde \Gamma^a_{(n+a)\ a} =  \tilde \Gamma^a_{(n+a) \ (n+a)} = -\tilde\Gamma^{n+a}_{a\ a} = \tilde\Gamma^{n+a}_{a \ (n+a)} = \frac{1}{2 \sqrt{p_a}} \qquad (a=1 \dots n).
	\eeq
Finally, the components of the curvature tensor of the canonical connection are 
	\beq
	\label{canCurv}
	\tilde R^i_{\ jkl} = \hat e_{(l)}\left(\tilde\Gamma^i_{jk} \right) - \hat e_{(k)}\left(\tilde\Gamma^i_{jl} \right) + \tilde\Gamma^m_{jk} \tilde\Gamma^i_{m l} - \tilde\Gamma^m_{jl} \tilde\Gamma^i_{mk} + \gamma^m_{kl} \tilde\Gamma^i_{jm}.
	\eeq
Using the definition of the canonical basis \eqref{canbasis} together with the the structure functions 
$\gamma^i_{jk}$ obtained from \eqref{strucF} and the expression for the non-vanishing connection symbols \eqref{nonzeroC} 
it can be directly verified that all the components of the Riemann tensor \eqref{canCurv} are identically zero [c.f. Appendix \ref{appendixC}].

 It is an interesting fact that the Levi-Civita and the canonical connections play a dual geometric role in the following sense: the former is the unique torsion-free and metric compatible connection whose curvature is non-trivial whereas the latter is the unique curvature-free and  \emph{fully} adapted connection with non-trivial torsion, providing us with two geometrically independent pictures of the same object. Indeed, the Ricci curvature of the Levi-Civita connection 
	\beq
	{\rm Ric}  =  - 2 n  \,\hat \theta^{(0)} \otimes \hat \theta^{(0)} + 2 \sum_{a=1}^n \left[\hat \theta^{(a)}_+ \otimes \hat \theta^{(a)}_+ - \hat \theta^{(a)}_- \otimes \hat \theta^{(a)}_- \right],
	\eeq
satisfies the property
	\beq
	{\rm Ric}(X,Y) = \lambda \eta(X) \eta(Y) + \nu G(X,Y)
	\eeq
where $\lambda = -(2n + 2)$ and $\nu = 2$. Thus, the Phase-Space is -- additionally -- an \emph{$\eta$-Einstein} para-Sasakian structure.

We conclude that the Thermodynamic Phase Space -- $(\mathcal{T}^{2n+1},\eta,\xi,\Phi,G)$ --  is a \emph{canonically flat} $\eta$-Einstein para-Sasakian manifold. 
 Hence,  
 it follows from the IVZ Theorem (c.f. Appendix \ref{appB}) that the Thermodynamic Phase Space is locally isomorphic to the hyperbolic Heisenberg group.
 However, the metrics on the hyperbolic Heisenberg group and on the Thermodynamic Phase Space are not exactly equivalent.
In fact, although the Ricci part of the curvature  is the same, the Weyl part   is different in the two cases. Furthermore, it can be verified that these metrics satisfy the field equations for the vacua of Einstein-Gauss-Bonnet gravity. This will be the subject of further investigation \cite{wip}.

%%%%%%%%%%%%%%%%%%%%%%%%%%%%%%%%%%%%%

\section{Closing Remarks}\label{sec5}

 %In fact, as we have shown, the geometry is a Sasakian geometry, a number of geometrical as well as
%topological consequences follow. For example, geodesics in the phase space have never been investigated in the literature, since it was
%not clear their physical meaning. However, it is known that for a Sasakian manifold, orbits generated by the Reeb vector field are always
%geodesics and they play an important role
%in the investigation of the symmetry group of the manifold itself. Accordingly, there is for example a classification of Sasakian manifolds into \emph{regular} and
%\emph{quasi-regular} or \emph{irregular} depending on whether such geodesics are all compact or at least one of them is not, respectively.

In this work,  we have presented a geometric formulation of the emergence of the macroscopic phase space of thermodynamics
based on the maximum entropy principle. This construction   unifies various aspects arising
from statistical mechanics, information theory and metric contact geometry. 
One important aspect of our formulation is that it can be generalized to the case of non-equilibrium systems by considering probability distributions different from the Gibbs one as reference in the relative entropy functional. %The  construction presented here is valid provided the  free entropy of the system under consideration has a 
%non-vanishing Hessian at the point in the control manifold where we want to define local equilibrium.  

%It is worth mentioning that the properties of the Thermodynamic Phase Space stemming from various geometrization programmes for thermodynamics  are not widely known. Indeed, the vast majority of  the work in this area has been confined
%to the investigation of the metric structures on the equilibrium manifold. This is mainly because the contact metric  structure of the Thermodynamic Phase 
%Space had no clear physical nor geometrical interpretation. This work  opens the possibility for new investigations in the geometrization of thermodynamics  
%through the study of the geometry of the Thermodynamic Phase Space.

We derived a number of useful properties of the phase space geometry. 
In particular, we showed that it is a para-Sasakian manifold defined by a metric of signature $(n+1,n)$ whose associated Levi-Civita connection satisfies  the defining property of an $\eta$-Einstein manifold.  Moreover, introducing another connection which is parallel with respect to the tensors defining the para-Sasakian structure and using the IVZ theorem [c.f. Appendix \ref{appB}] we have shown that this manifold is locally isomorphic to the hyperbolic Heisenberg group.

The relationship with the Heisenberg group is not surprising since, by construction, the observables we considered correspond to mean values of random variables defined on a suitable microscopic space of events. Such a set up is sufficiently general in the sense that the data measured in an experiment can never be known with full accuracy; we can never achieve complete knowledge of the defining variables of a system. This is the starting point for the general problem of model selection in statistical inference theory.  In this setting,  the maximum entropy principle is a criterion for selecting the `optimal' probability distribution  consistent with the given data: the one which is least informative about anything else not contained in the data set.  Therefore, there is an intrinsic uncertainty encoded in the very definition of the Thermodynamic Phase Space which manifests itself by restricting the geometry to be the hyperbolic Heisenberg group. %The hyperbolic nature of the Thermodynamic Phase Space can be directly related to chaotic motion, which is a necessary ingredient in order to obtain 
%correct statistical mechanical calculations, although often it is not explicitly mentioned [CITATION].

Let us close this work by pointing some future directions. Firstly, the signature of the metric of the Thermodynamic Phase Space allows for the existence of null directions providing us with a cone structure similar to the one {  present} in Relativity. The existence of such a cone structure is a signal that there may be a way of characterizing correlations between states (points in the Thermodynamic Phase Space) in a geometric way analogous to that of points in a space-time. Secondly,  the hyperbolic nature of the Thermodynamic Phase Space remains to be studied, a possible direction can be directly related to chaotic motion, which is a necessary ingredient in order to obtain correct statistical mechanical calculations. %, although often it is not explicitly mentioned [CITATION]. 
Finally, noting that the Thermodynamic Phase Space has constant Levi-Civita scalar curvature suggests that it is highly symmetric. Indeed, the Levi-Civita connection satisfies the defining condition of an $\eta$-Einstein manifold, a slight modification of an Einstein manifold. In gravitational theories, an important example are the dS and AdS solutions for the vacuum Einstein Field Equations with a cosmological constant. It can be directly verified that the metric presented in this work is a solution for the vacuum of Einstein-Gauss-Bonnet theory of gravity analogous to the de Sitter solution. This is a remarkable connection between information geometry and gravity which deserves further exploration \cite{wip}.

%.    It follows that the Heisenberg commutation relations
%characterizing quantum mechanics 
%are already contained in the macroscopic thermodynamic -- observable -- phase space, even though we have started from classical statistical mechanics.
%We believe that this is due to the fact that statistical mechanics introduces an ignorance over the full microscopic dynamics of the system.
%This ignorance is encoded in the fact that one macroscopically can only see mean values of determined `observable' quantities and is maximized
%in order to obtain the equilibrium distribution, which is the one that gives the maximum entropy -- i.e. the maximum of ignorance -- and is compatible with the
%observed data.
%It is remarkable that the introduction of ignorance into classical statistical mechanics leads to Heisenberg dispersion relations.
%Moreover, the phase space is also hyperbolic. This type of geometry is directly related to chaotic motion, which is a necessary ingredient in order to obtain 
%correct statistical mechanical calculations, although often it is not explicitly remarked.
%We wish to explore further all these interesting geometric properties of the macroscopic phase space of thermodynamics in future works.
%We believe that their investigations can shed new lights over the relationships between information and geometry, as well as microscopic and macroscopic observations.
%In this sense they can help to construct a geometric theory of ignorance. 

\section*{Acknowledgements}
The authors are thankful to F Nettel, H Quevedo, G Arciniega and C Gruber for insightful comments and suggestions. 
AB wants to express his gratitude  
to the A. Della Riccia Foundation (Florence, Italy) for financial support. 
CSLM acknowledges financial support from DGAPA-UNAM (postdoctoral fellowship).

\appendix

\section{Calculation of the canonical curvature}
\label{appendixC}
In this appendix we compute all the non-trivial components of the  Riemann tensor of the canonical connection, i.e.
	\beq
	\label{canCurvAp}
	\tilde R^i_{\ jkl} = \hat e_{(l)}\left(\tilde\Gamma^i_{jk} \right) - \hat e_{(k)}\left(\tilde\Gamma^i_{jl} \right) + \tilde\Gamma^m_{jk} \tilde\Gamma^i_{m l} - \tilde\Gamma^m_{jl} \tilde\Gamma^i_{mk} + \gamma^m_{kl} \tilde\Gamma^i_{jm}.
	\eeq
 Firstly, the non-vanishing structure functions of the canonical basis  [c.f. equation \eqref{strucF}] are given by
	\beq
	\gamma^0_{a\ (n+a)} = 2 \quad  \text{and} \quad \gamma^a_{a \ (n+a)} = \gamma^{n+a}_{a \ (n+a)}  =  -\frac{1}{2\sqrt{p_a}}.
	\eeq
Note that these coefficients are anti-symmetric in their lower indices. Notice as well that -- from \eqref{nonzeroC} -- the curvature tensor does not have $i,j,k,l = 0$. Thus, let us split the calculation in its eight different non-trivial possibilities. In the following, we do not use Einstein sum convention except for the label `m'.
	\begin{enumerate}
		 \item $i=a$, $j=a$ and $k=a$ case.
			\begin{align}
			\tilde R^a_{\ a a l} 	&= \hat e_{(l)} \tGamma^a_{a a} - \hat e^+_{(a)} \tGamma^a_{a l} + \tGamma^m_{aa} \tGamma^a_{ml} - \tGamma^m_{al}\tGamma^a_{ma} + \gamma^m_{a l} \tGamma^a_{a m}\nonumber\\ 
									&= \tGamma^{n+a}_{a \ a} \tGamma^a_{(n+a)\ l} - \tGamma^{n+a}_{a \ l}\tGamma^a_{(n+a)\ a}\nonumber\\
									&=\left\{\begin{array}{lr}
													\left(-\frac{1}{\sqrt{p_a}} \right)\left(-\frac{1}{\sqrt{p_a}} \right) -\left(-\frac{1}{\sqrt{p_a}} \right)\left(-\frac{1}{\sqrt{p_a}}\right), & l=a\\
													\left(-\frac{1}{\sqrt{p_a}} \right)\left(\frac{1}{\sqrt{p_a}} \right) -\left(\frac{1}{\sqrt{p_a}} \right)\left(-\frac{1}{\sqrt{p_a}}\right), & l=n+a
													\end{array}\right. \nonumber\\
									& = 0.
			\end{align}
			
			\item $i=a$, $j=a$ and $k=n+a$ case.
				\begin{align}
				\tilde R^a_{\ a (n+a) l} & = \hat e_{(l)}\tGamma^a_{a\ (n+a)} - \hat e_{(n+a)} \tGamma^a_{a l} + \tGamma^m_{a \ (n+a)} \tGamma^a_{m l} - \tGamma^m_{a l} \tGamma^a_{m \ (n+a)} + \gamma^m_{(n+a) l} \Gamma^a_{a \ (n+a)}\nonumber\\
											& = \tGamma^{n+a}_{a\ (n+a)} \tGamma^a_{(n+a) \ l} - \left[ \tGamma^{n+a}_{a \ l} \tGamma^a_{(n+a) (n+a)} + \tGamma^a_{al} \tGamma^a_{a \ (n+a)}\right]\nonumber\\
											&= \left\{\begin{array}{lr}
																\left(\frac{1}{\sqrt{p_a}} \right)\left(-\frac{1}{\sqrt{p_a}} \right) -\left(-\frac{1}{\sqrt{p_a}} \right)\left(\frac{1}{\sqrt{p_a}}\right), & l=a\\
																\left(\frac{1}{\sqrt{p_a}} \right)\left(\frac{1}{\sqrt{p_a}} \right) -\left(\frac{1}{\sqrt{p_a}} \right)\left(\frac{1}{\sqrt{p_a}}\right), & l=n+a\\
																\end{array}\right. \nonumber\\
											&= 0.
				\end{align}
				
				\item $i=a$, $j=n+a$ and $k=a$ case.
					\begin{align}
					\tilde R^a_{\ (n+a)a l} 		&= \hat e_{(l)}\tGamma^a_{(n+a)a} - \hat e_{(a)}\tGamma^a_{(n+a)l} + \tGamma^m_{(n+a)a} \tGamma^a_{ml} - \tGamma^m_{(n+a)l}\tGamma^a_{ma}+\gamma^m_{al}\tGamma^a_{(n+a) m}\nonumber\\
													&= \hat e_{(l)}\left(-\frac{1}{2 \sqrt{p_a}} \right) - \hat e_{(a)} \tGamma^a_{(n+a)l}\nonumber\\
													& =\left\{\begin{array}{lr}
																\frac{\partial}{\partial p_a} \left(-\frac{1}{2 \sqrt{p_a}} \right) - \frac{\partial}{\partial p_a}\left(\frac{1}{2\sqrt{p_a}} \right) 	& l=a \\
																-\frac{\partial}{\partial p_a}\left(-\frac{1}{2 \sqrt{p_a}} \right) - \frac{\partial}{\partial p_a}\left(\frac{1}{2\sqrt{p_a}} \right)		& l=n+a\\
																\end{array}\right.\nonumber\\
													&= 0.
					\end{align}
					
				\item $i=a$, $j=n+a$ and $k=n+a$ case.
					\begin{align}
						\tilde R^a_{\ (n+a)(n+a) l}	&= \hat e_{(l)} \tGamma^a_{(n+a)(n+a)} - \hat e_{(n+a)} \tGamma^a_{(n+a)l} + \tGamma^m_{(n+a)(n+a)}\tGamma^a_{m l} - \tGamma^m_{(n+a)l}\tGamma^a_{m(n+a)} + \gamma^m_{(n+a) l} \tGamma^a_{(n+a) m}\nonumber\\
															&=\hat e_{(l)}\left(\frac{1}{2\sqrt{p_a}} \right) + \frac{\partial}{\partial p_a} \left(\tGamma^a_{(n+a)l} \right) + \gamma^a_{(n+a)l} \tGamma^a_{(n+a) a} + \gamma^{(n+a)}_{(n+a)l}\tGamma^a_{(n+a)(n+a)}\nonumber\\
															&=\left\{\begin{array}{lr}
																			\frac{\partial}{\partial p_a} \left(\frac{1}{2\sqrt{p_a}} \right) - \frac{\partial}{\partial p_a} \left(\frac{1}{2\sqrt{p_a}} \right) + \frac{1}{2\sqrt{p_a}}\left(- \frac{1}{2\sqrt{p_a}} \right) + \frac{1}{2\sqrt{p_a}}\left(\frac{1}{2\sqrt{p_a}} \right), 	& l=a \\
																			-\frac{\partial}{\partial p_a} \left(\frac{1}{2\sqrt{p_a}} \right) + \frac{\partial}{\partial p_a} \left(\frac{1}{2\sqrt{p_a}} \right) , 	& l=n+a\\
																			\end{array}\right.\nonumber\\
															&=0.
					\end{align}
					
				\item $i=n+a$, $j=a$ and $k=a$ case.
					\begin{align}
						\tilde R^{n+a}_{\ a a l} 	&= \hat e_{(l)}\tGamma^{(n+a)}_{a a} - \hat e_{(a)} \tGamma^{(n+a)}_{a l} + \tGamma^{m}_{a a} \tGamma^{(n+a)}_{m l} - \tGamma^{m}_{a l}\tGamma^{(n+a)}_{m a} + \gamma^m_{a l} \tGamma^{(n+a)}_{a m}\nonumber\\
														& = \hat e_{(l)} \left(-\frac{1}{2\sqrt{p_a}} \right) - \hat e_{(a)}\tGamma^{(n+a)}_{a l}\nonumber\\
														& = \left\{\begin{array}{lr}
																			\frac{\partial}{\partial p_a} \left(-\frac{1}{2\sqrt{p_a}} \right) - \frac{\partial}{\partial p_a}\left(-\frac{1}{2\sqrt{p_a}} \right), 	& l=a\\
																			\frac{\partial}{\partial p_a} \left(\frac{1}{2\sqrt{p_a}} \right) - \frac{\partial}{\partial p_a}\left(\frac{1}{2\sqrt{p_a}} \right), 		& l=n+a\\
																			\end{array}\right.\nonumber\\
														& = 0.
					\end{align}
					
				\item $i=n+a$, $j=a$ and $k=n+a$ case.
					\begin{align}
						\tilde R^{n+a}_{\ a (n+a) l} 	& = \hat e_{(l)}\tGamma^{(n+a)}_{a (n+a)} - \hat e_{(n+a)} \tGamma^{(n+a)}_{a l} + \tGamma^m_{a (n+a)} \tGamma^{(n+a)}_{m l} - \tGamma^m_{a l}\tGamma^{(n+a)}_{m (n+a)} + \gamma^m_{(n+a) l} \tGamma^{(n+a)}_{a m}\nonumber\\
															& = \hat e_{(l)} \left(\frac{1}{2\sqrt{p_a}} \right) + \frac{\partial}{\partial p_a}  \tGamma^{(n+a)}_{al}\nonumber\\
															& = \left\{\begin{array}{lr}
																			\frac{\partial}{\partial p_a} \left(\frac{1}{2\sqrt{p_a}} \right) + \frac{\partial}{\partial p_a} \left(- \frac{1}{2\sqrt{p_a}} \right), 	& l=a \\
																			-\frac{\partial}{\partial p_a} \left( \frac{1}{2\sqrt{p_a}}\right) + \frac{\partial}{\partial p_a}\left(\frac{1}{2\sqrt{p_a}} \right),		& l=n+a\\
																			\end{array}\right.\nonumber\\
															& = 0.
					\end{align}
					
				\item $i=n+a$, $j=n+a$ and $k=a$ case.
					\begin{align}
						\tilde R^{n+a}_{\ (n+a) a l}	& = \hat e_{(l)}\tGamma^{(n+a)}_{(n+a) a} - \hat e_{(a)}\tGamma^{(n+a)}_{(n+a) l} + \tGamma^{m}_{(n+a) a} \tGamma^{(n+a)}_{m l} - \tGamma^{m}_{(n+a) l} \tGamma^{(n+a)}_{m a} + \gamma^{m}_{a l}\tGamma^{(n+a)}_{(n+a) m} \nonumber\\
															& = \tGamma^{a}_{(n+a)a} \tGamma^{(n+a)}_{a l} - \tGamma^{a}_{(n+a) l }\tGamma^{(n+a)}_{aa}\nonumber\\
															& = \left\{ \begin{array}{lr}
																					\left(-\frac{1}{2\sqrt{p_a}} \right)\left(-\frac{1}{2\sqrt{p_a}} \right) - \left(-\frac{1}{2\sqrt{p_a}} \right)\left(-\frac{1}{2\sqrt{p_a}} \right), 	& l=a \\
																					\left(-\frac{1}{2\sqrt{p_a}} \right)\left(\frac{1}{2\sqrt{p_a}} \right) - \left(\frac{1}{2\sqrt{p_a}} \right)\left(-\frac{1}{2\sqrt{p_a}} \right),		& l=n+a\\
																				\end{array}\right.\nonumber\\
															& = 0.
					\end{align}
					
				\item $i=n+a$, $j=n+a$ and $k=n+a$ case.
					\begin{align}
						\tilde R^{n+a}_{\ (n+a) (n+a) l} 	& = \hat e_{(l)}\tGamma^{(n+a)}_{(n+a)(n+a)} - \hat e_{(n+a)} \tGamma^{(n+a)}_{(n+a) l} + \tGamma^{m}_{(n+a)(n+a)}\tGamma^{(n+a)}_{m l} - \tGamma^{m}_{(n+a) l}\tGamma^{(n+a)}_{m (n+a)} + \gamma^{m}_{(n+a) l} \tGamma^{(n+a)}_{(n+a) m}\nonumber\\
																	& = \tGamma^{a}_{(n+a)(n+a)}\tGamma^{(n+a)}_{a l} - \tGamma^{a}_{(n+a) l}\tGamma^{(n+a)}_{a (n+a)}\nonumber\\
																	& = \left\{ \begin{array}{lr}
																						\left(\frac{1}{2\sqrt{p_a}} \right)\left(-\frac{1}{2\sqrt{p_a}} \right) - \left(-\frac{1}{2\sqrt{p_a}} \right)\left(\frac{1}{2\sqrt{p_a}} \right),	& l=a\\
																						\left(\frac{1}{2\sqrt{p_a}} \right)\left(\frac{1}{2\sqrt{p_a}} \right) - \left(\frac{1}{2\sqrt{p_a}} \right)\left(\frac{1}{2\sqrt{p_a}} \right),		& l=n+a\\
																						\end{array}\right.\nonumber\\
																	& = 0.
					\end{align}
				
	\end{enumerate}

Therefore, the Riemann tensor of the canonical connection is identically zero.

\section{The hyperbolic Heinseberg group}
\label{appB}
In this appendix we give the definition of the hyperbolic Heisenberg group as in \cite{IVZ} and we state the Theorem 4.2 in 
the same reference.

Let us introduce a manifold $G(\mathbb P)=\mathbb R^{2n}\times \mathbb R$ with the group law
	\beq\label{grouplaw}
	(p'',t'')=(p',t')\cdot(p,t)=\left(p'+p,t'+t-\sum_{k=1}^{n}(u_{k}'\,v_{k}-v_{k}'\,u_{k})\right)\,,
	\eeq
with $p=(u_{1},v_{1},\dots,u_{n},v_{n})$, $p'=(u_{1}',v_{1}',\dots,u_{n}',v_{n}')$ and $t,t' \in \mathbb R$.
The contact structure is given by the $1$-form 
	\beq
	\label{tildeTheta}
	\tilde\Theta=-\frac{1}{2}\d t - \sum_{k=1}^{n}(u_{k}\d v_{k}-v_{k}\d u_{k})
	\eeq
and therefore the vector fields
	\beq\label{orthbasisH}
	\tilde\xi=2\frac{\partial}{\partial t}, \quad U_{k}=\frac{\partial}{\partial u_{k}}-2v_{k}\frac{\partial}{\partial t}, \quad V_{k}=\frac{\partial}{\partial v_{k}}+2u_{k}\frac{\partial}{\partial t}
	\eeq
form the canonical basis for the tangent space, with the vectors $U_{k}$ and $V_{k}$ spanning the horizontal distribution $\tilde{\mathcal{D}}={\rm ker}\tilde\Theta$ and $\tilde\xi$ spanning
the vertical direction, as in \eqref{splitting}.
One can give such manifold a para-contact structure $\tilde\Phi$ defined by the following rules
	\beq\label{paracontactH}
	\tilde\Phi\,\tilde\xi=0,\quad \tilde\Phi\, U_{k}=V_{k}, \quad \tilde\Phi\, V_{k}=U_{k}\,.
	\eeq
Finally, one can define on $G(\mathbb P)$ a metric such that
	\beq\label{metricH}
	\tilde G(\tilde\xi,\tilde\xi)=1,\quad \tilde G(U_{k},U_{k})=1, \quad \tilde G(V_{k},V_{k})=-1\,,
	\eeq
so that the canonical basis \eqref{orthbasisH} turns out to be the orthonormal basis for this metric.
Note that this metric differs from the standard Sasaki metric on the Heisenberg group in the signature, while the group laws are the same.
The structure $(G(\mathbb P), \tilde \Theta, \tilde\xi, \tilde\Phi, \tilde G)$ is the \emph{hyperbolic Heisenberg group}. 
It is an example of an integrable para-contact hermitian structure with flat canonical connection.
Moreover, the following theorem from \cite{IVZ} states that locally it is the only such example.
	\begin{theo}[IVZ Theorem] Let $\left(M,\eta,\xi,\Phi,G\right)$ be an integrable para-contact hermitian manifold of dimension $2n+1$. 
	\begin{itemize}
	\item[i)] If $n>1$, then  $\left(M,\eta,\xi,\Phi,G\right)$ is locally isomorphic to the hyperbolic Heisenberg group if and only if the canonical 
	connection \eqref{canonicalconn} has vanishing horizontal curvature.
	\item[ii)] If $n=1$, then   $\left(M,\eta,\xi,\Phi,G\right)$ is locally isomorphic to the $3$-dimensional hyperbolic Heisenberg group 
	if and only if the canonical 
	connection \eqref{canonicalconn} has vanishing horizontal curvature and zero torsion.
	\end{itemize}
	\end{theo}

\end{document}